\documentclass{ws-p9-75x6-50}

\begin{document}

\title{Transport in quantum wires with impurities at finite temperature}

\author{T. Kleimann, M. Sassetti}

\address{Dipartimento di Fisica, INFM, Universit\`{a} di Genova, 16146
  Genova, Italy\\E-mail: kleimann@fisica.unige.it,
  sassetti@fisica.unige.it}

\author{B. Kramer}

\address{I. Institut f{\"ur} Theoretische Physik, Universit{\"a}t
  Hamburg,\\ Jungiusstra\ss{}e 9, 20355 Hamburg, Germany\\E-mail:
  kramer@physnet.uni-hamburg.de}


\maketitle

\abstracts{The temperature dependence of Coulomb blockade peaks of a
  one dimensional quantum dot is calculated. The Coulomb interaction
  is treated microscopically using the Luttinger liquid model. The
  electron interaction is assumed to be non-homogeneous with a maximum
  strength near the quantum dot. The conductance peaks show
  non-analytic power law behaviour induced by the interaction. It is
  shown that there is a crossover in the power law which is related to
  the inhomogeneity of the interaction.}

\section{Introduction}
One dimensional quantum dots can be formed between two impurities in a
semiconductor quantum wire when the electron density is sufficiently
low \cite{aus}. Transport measurements at such dots reflect the
properties of the interaction between the electrons. Especially, the
temperature power-law dependence of the Coulomb blockade conductance
peaks is given by the interaction strength. In the present paper, we
show that if the (short-ranged) interaction is non-homogeneous, i.e.
\begin{equation}
  \label{eq:1}
V(x,y)= \left[ V_0 +\varphi(x) \right] \delta (x-y),   
\end{equation}
there is a crossover in the temperature-power law between the
interaction strength near the quantum dot ($|x| \ll b$, maximum of
$\varphi(x)$) for high temperature, and the interaction strength far
away from the dot ($|x|\gg b$, $\varphi(x)\approx 0$) at low
temperature.

\section{Inhomogeneous Luttinger Liquid}
Using the Bosonisation method the excitations in the spinless one
dimensional (1D) interacting electron system are described by charge
density waves in terms of the conjugate Boson fields \cite{voit},
$[\vartheta(x),\Pi(x')]=i \delta(x-x')$. The Hamiltonian
is
\begin{equation}
H=\frac{\hbar v_{\rm F}}{2} 
\int {\rm d}x \left\{ \Pi^2(x) + 
\frac{1}{g_0^2}[\partial_x \vartheta(x)]^2 \right\} + 
\frac{1}{2 \pi} \int {\rm d}x \varphi(x) 
\left[ \partial_x \vartheta(x) \right]^2
\end{equation}
with $v_{\rm F}$ the Fermi velocity. The fields are related to the
electron density $\rho$ via $\partial_x \vartheta(x) = \sqrt{\pi}
\left[ \rho(x) -\rho_0\right]$ where $\rho_0=k_{\rm F}/\pi$ is the
mean density. The parameter $g_0=[1+V_0/\pi \hbar v_{\rm F}]^{-1/2}$
is the interaction constant that arises from the asymptotic constant
part of the interaction for $|x|\to \infty$. The second term is due to
the inhomogeneity treated as a perturbation. It is considered here
only in lowest order.  The temperature Green function
$G(x,x';\tau,\tau')=\langle T_\tau
[\vartheta(x,\tau)\vartheta(x',\tau')] \rangle$, where $T_\tau$ stands
for imaginary time ordering, is
\begin{equation}
  \label{eq:greens}
    G_{\omega_n}(x, x') 
  = G^0_{\omega_n}(x-x') +\frac{1}{\pi} \int {\rm d}y \; 
G^0_{\omega_n}(x-y) \left[ \partial_y \varphi(y) \partial_y \right]  
G^0_{\omega_n}(y-x').
\end{equation}
The unperturbed Green function is $G^0_{\omega_n}=(g_0/2\hbar
\omega_n) \exp{\{ -g_0|\omega_n(x-x')|/v_{\rm F}\}}$ at the Matsubara
frequencies $\omega_n=2\pi n/\beta\hbar$ whith the inverse
temperature $\beta=1/k_{\rm B}T$.

The transport properties are related to the propagator
(\ref{eq:greens}). For instance, the nonlocal DC-conductivity is
$\sigma_{\omega_n\rightarrow 0}(x,x')=\lim_{\omega_n\rightarrow 0}
e^2\omega_n G_{\omega_n}(x,x')/\pi$. This leads to the conductance
quantum $g_0 e^2/h$. Hence, the conductance of a pure quantum wire is
given by the interaction parameter for $|x|\to\infty$, $g_0$. It does
not depend on the interaction within the wire \cite{rencond}.

\section{Double Barrier}
In the experiment, the electron density of the 1D electron system is
lowered such that two maxima of the impurity potential are higher
than the Fermi energy. Thus, a 1D quantum dot is defined between the
maxima. We model this by assuming $V_{\rm d}(x)=V_{\rm d} \delta (x-x_i)$
where $x_i=x_{\rm d}$, $x_{\rm d}+a$. The corresponding Hamiltonian is
$H_{\rm d}= V_{\rm d} \sum_i \cos \left[2k_{\rm F} x_i + 2\sqrt{\pi}
  \vartheta(x_i) \right]$. 

For the conductance we need to calculate the effective action
corresponding to the Hamiltonian. This is done with the imaginary time
path integral method. The variables at $x\neq x_i$ are traced out.
Eventually one obtains the effective action in the variables
$\theta_\pm=(\theta(x_{\rm d})\pm \theta(x_{\rm d}+a))/2$
\begin{equation}
  \label{eq:effac}
   S_{\rm eff}= \frac{1}{2 \hbar\beta} \sum_{n, \pm} \theta_\pm(\omega_n) 
K_\pm(\omega_n) \theta_\pm(\omega_n) + S_{\rm d}[\theta_+, \theta_-],
\end{equation}
where $S_{\rm d}$ corresponds to $H_{\rm d}$.

The kernels depend directly on the one- and two-point functions of the
inhomogeneous Luttinger system and can be written as $K_\pm(\omega_n)
=\pi/2 [G_{\omega_n}(x_{\rm d},x_{\rm d}) \pm G_{\omega_n}(x_{\rm d},
x_{\rm d}+a)]$. We assumed the distance between the barriers, $a$,
beeing much smaller than the extension $b$ of the inhomogeneity
potential, $a \ll b$ such that the interaction varies only very slowly
within the dot. The interaction parameter at the position of the
double barrier is $g(x_{\rm d})=[1+ (V_0+ \varphi(x_{\rm d}))/\pi
\hbar v_{\rm F}]^{-1/2}$. The charging energy of the dot then reads
\begin{equation}
E_{\rm c} = K_-(\omega_n \downarrow 0) = 
\frac{\pi \hbar v_{\rm F}}{2 a g^2(x_{\rm d})}.
\end{equation}
Thus, the charging energy is a local quantity which depends
on the length of the dot and the repulsion strength in the
dot region through $g(x_{\rm d})$. 

The sequential tunneling rates are obtained in the limit of high
barriers \cite{weiss}
\begin{equation}
  \Gamma(E)= \left(\frac{\Delta}{2}\right)^2 
\int_{-\infty}^{\infty} {\rm d}t \; e^{iEt/\hbar-W(t)}.
\end{equation}
The energies $E$ correspond to forward and backward tunneling through
the left and right barrier respectively. The tunneling amplitude
$\Delta$ is related to $V_{\rm d}$ via the WKB-method. The kernel
$W(t)$ contains the information about the excitations in the leads and
the excited states in the dot.  By using the above
Green function,
\begin{equation}
  \label{eq:kern}
  W(t)=\int_{-\infty}^\infty \frac{{\rm d}\omega}{2\hbar \omega^2}
\,{\rm Im} \frac{1}
{G_{\omega_n= -i\omega}(x_{\rm d},x_{\rm d})} 
 \, \Big[1 + \epsilon \sum_n 
\delta (\hbar \omega - n \epsilon) \Big] 
\frac{1-e^{-i\omega t}}{1-e^{-\hbar \omega \beta}} 
\;e^{-\omega/\omega_{\rm c}}
\end{equation}
where $\omega_{\rm c}$ is the high frequency cutoff. The energy $\epsilon
\equiv \pi \hbar v_{\rm F}/a g(x_{\rm d})=2 g(x_{\rm d}) E_{\rm c}$ is
the discrete level spacing of the plasmon states in the quantum dot.
As the charging energy, this depends only on the interaction near the
quantum dot, i.e. the local interaction strength $g(x_{\rm d})$. This
holds for arbitrary shapes of $\varphi(x)$.

\section{Linear Transport}
\begin{figure}[t]
\epsfxsize=11.1cm 
\epsfbox{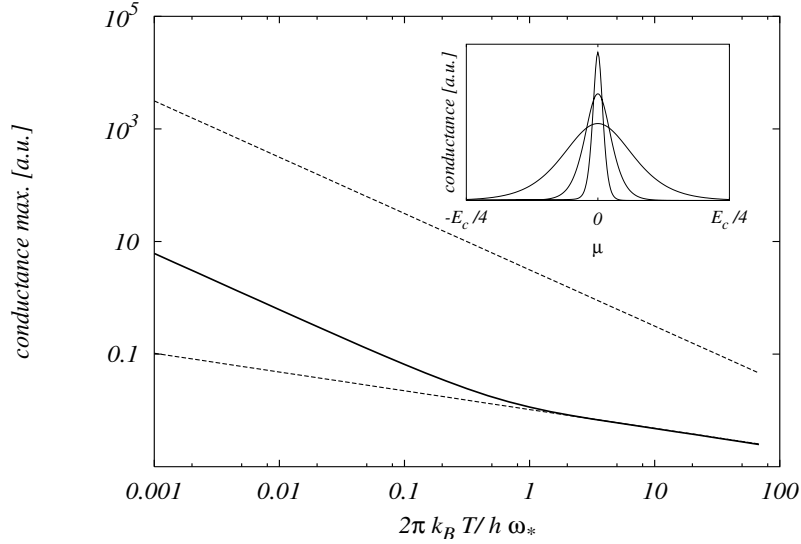} 

\caption{\label{fig:peaks}Logarithmic dependence of the 
  conductance maximum on temperature $T$ for $g_0=1.0$ and $g(x_{\rm
    d})=0.6$ (full curve). Parameters are $\omega_\ast=v_{\rm F}/g_0
  b$ crossover frequency; dot position in the centre of the
  inhomogeneity $x_{\rm d}=0$; ratio between dot length and extension
  of the inhomogeneity $a/b=0.1$. Straight dotted lines: homogeneous
  limits for $g(x)=g_0$ and $g(x)=g(x_{\rm d})$. Inset: Coulomb
  blockade conductance peaks for $k_{\rm B}T/\hbar \omega_\ast=$0.05,
  0.1, 0.4 (top to bottom at $\mu =0$); $E_{\rm c}$ charging energy; peaks
  corresponding to $g_0$ and $g(x_{\rm d})$ are out of the scale of
  the inset.}
\end{figure}

In the limit of linear transport the chemical potentials in the left
and right leads and the dot are aligned. Then Coulomb blockade is
relaxed and the conductance versus the gate voltage shows a peak. For
sequential tunneling we can use the master equation method
\cite{grade} for calculating \cite{braggio} the conductance for
$k_{\rm B}T \ll \epsilon$
\begin{equation}
  \label{eq:cond}
  {\cal G}= \frac{e^2 \beta}{4}
\frac{e^{-\beta \mu/2}}{\cosh \beta \mu/2} \Gamma(\mu).
\end{equation}
Here $\mu$ is the distance from the resonance energy. In the limit
$\varphi(x)\equiv 0$ one can obtain the temperature behaviour of the
conductance maximum analytically: ${\cal G}^{\rm max} \propto (\beta
\hbar \omega_{\rm c})^{2-1/g_0}$. In general one needs to calculate ${\cal
  G}$ numerically. Figure \ref{fig:peaks} shows the peak shape and the
temperature behaviour of the maximum for
$\varphi(x)=V_1/(1+4x^2/b^2)$.
Contrary to the na\"{\i}ve guess, namely that the exponent contained
simply the local interaction parameter $g(x_{\rm d})$, we observe a
crossover from an effective $g_{\rm eff}$ resembling $g(x_{\rm d})$ at
high temperatures towards $g_0$ at low temperatures. The crossover
temperature is related to the extension of the inhomogeneity $b$.
Thus, the temperature power law can be governed by any value of
$g_{\rm eff}$ between the two asymptotic limits. Our finding leads to
a better understanding of the controversy between the experimentally
determined interaction parameter in nonlinear and linear transport
through a 1D quantum dot device \cite{aus,kleim}. The nonlinear
transport experiment probes the local interaction at the position of
the dot ($g(x_{\rm d})<g_0$) in terms of level spacing and charging
energy. However, the examination of the conductance peaks involves the
entire Luttinger liquid system with its excitations and the
inhomogeneity. Hence an effective $g$-parameter depending on the
temperature range of the experiment can in principle be extracted from
the power law behaviour of the peak maximum.

Preliminary results suggest that for a considerable inhomogeneity the
temperature behaviour could even be reversed during the crossover from
$g(x_{\rm d})$ to $g_0$ and thus indicating almost arbitrary effective
interaction parameters extracted from experiments in such temperature
regimes.

\section{Conclusion}
For the example of the temperature dependence of the Coulomb blockade
conductance peaks of a 1D quantum dot, we have shown that
inhomogeneity effects in a Luttinger system can lead to considerable
changes in the non-analytic power law. Depending on the temperature
region, different exponents can occur that reflect the strength of the
interaction in different parts of the system. Our results indicate
that deducing Luttinger liquid parameters from different sets of
experimental data might lead to contradicting numerical values. In
particular, the above model opens a possibility to explain certain 
discrepancies in the parameters obtained from cleaved-edge overgrowth
quantum dots \cite{aus,kleim}.

\end{document}